\documentclass[pra,twocolumn,superscriptaddress,showpacs,amsmath,amssymb]{revtex4-1}
\usepackage{graphicx}
\usepackage{bm}
\usepackage{amssymb}
\usepackage{amsmath}
\usepackage{dcolumn}
\usepackage{color}
\usepackage{natbib}
\def\U#1{{\rm #1}}

\newcommand{\expect}[1]{\langle #1 \rangle} 
\newcommand{\normal}[1]{\mbox{$\langle :$} #1\mbox{$: \rangle$}} 
\def\H{{\rm H}}
\def\V{{\rm V}}

\def\00{\H\V}
\def\11{\V\H}

\begin{document}
\title{
A high visibility Hong-Ou-Mandel interference via a time-resolved coincidence measurement
}

\author{Yoshiaki~Tsujimoto}
\affiliation{Graduate School of Engineering Science, Osaka University,
Toyonaka, Osaka 560-8531, Japan}
\author{Yukihiro~Sugiura}
\affiliation{Graduate School of Engineering Science, Osaka University,
Toyonaka, Osaka 560-8531, Japan}
\author{Motoki~Tanaka}
\affiliation{Graduate School of Engineering Science, Osaka University,
Toyonaka, Osaka 560-8531, Japan}
\author{Rikizo~Ikuta}
\affiliation{Graduate School of Engineering Science, Osaka University,
Toyonaka, Osaka 560-8531, Japan}
\author{Shigehito Miki}
\affiliation{Advanced ICT Research Institute, National Institute of 
Information and Communications Technology (NICT), Kobe 651-2492, Japan}
\author{Taro Yamashita}
\affiliation{Advanced ICT Research Institute, National Institute of 
Information and Communications Technology (NICT), Kobe 651-2492, Japan}
\author{Hirotaka Terai}
\affiliation{Advanced ICT Research Institute, National Institute of 
Information and Communications Technology (NICT), Kobe 651-2492, Japan}
\author{Mikio Fujiwara}
\affiliation{Advanced ICT Research Institute, National Institute of 
Information and Communications Technology (NICT), Koganei, Tokyo 
184-8795, Japan}
\author{Takashi~Yamamoto}
\affiliation{Graduate School of Engineering Science, Osaka University,
Toyonaka, Osaka 560-8531, Japan}
\author{Masato~Koashi}
\affiliation{Photon Science Center, The University of Tokyo, 
Bunkyo-ku, 113-8656, Japan}
\author{Masahide Sasaki}
\affiliation{Advanced ICT Research Institute, National Institute of 
Information and Communications Technology (NICT), Koganei, Tokyo 
184-8795, Japan}
\author{Nobuyuki~Imoto}
\affiliation{Graduate School of Engineering Science, Osaka University,
Toyonaka, Osaka 560-8531, Japan}

\begin{abstract}
We report on the observation of a high visibility Hong-Ou-Mandel interference of 
two heralded photons emitted from a spontaneous parametric down conversion~(SPDC) 
pumped by continuous-wave~(cw) light. 
A non-degenerate photon pair at 1541~nm and 1580~nm is generated by cw-pumped SPDC 
through a periodically poled lithium niobate waveguide. 
The heralded single photon at 1541~nm is prepared by the detection of the photon at 
1580~nm. We performed the experiment of the Hong-Ou-Mandel interference between heralded single photons in separated
time bins and observed a high visibility interference. All detectors we used are 
superconducting nanowire single-photon detectors and an overall temporal 
resolution of the photon detection is estimated as 85 ps, which is sufficiently 
shorter than the coherence time of the heralded photons.
\end{abstract}

\maketitle
\section{Introduction}
Generation, manipulation and measurement of photonic quantum states 
are prerequisites for many applications of quantum information processing. 
A high visibility Hong-Ou-Mandel~(HOM) interference~\cite{PhysRevLett.59.2044} of 
photons generated by independent photon sources is an important issue for fulfilling such 
requirements~\cite{RevModPhys.84.777, ma2012quantum, Carolan14082015, tillmann2013experimental}. 
So far, many of such experiments have been performed 
with photons generated from spontaneous parametric down conversion (SPDC) 
pumped by ultrafast pulsed lasers~\cite{PhysRevA.67.022301, PhysRevLett.96.110501, PhysRevA.79.040302, PhysRevA.81.021801, Tanida:12,
mcmillan2013two, Tsujimoto:15}. 
In such experiments, the timing synchronization of the photons from the 
separated sources is achieved by the spectral filtering of the photons such 
that their coherence time is longer than that of the pump light 
and the adjustment of the optical path lengths 
for a high visibility interference. 
On the other hand, if timing resolutions of the photon detectors 
is sufficiently shorter than the coherence time of the photons, 
the temporal indistinguishability of the photons can be guaranteed 
by the coincidence detection with such a high timing resolution~\cite{PhysRevLett.71.4287}. 
In this case, even when the photons are generated by a cw pump laser, 
post analysis of the timings of the photon detections allows us to select 
photons having a sufficient temporal overlap. One of the advantages of this method 
is that there is no need to perform the active synchronizations of the photon sources. 
Such an experiment has been demonstrated in Ref.~\cite{halder2007entangling, 1367-2630-10-2-023027, PhysRevA.80.042321, 6985715}. 
In Ref.~\cite{halder2007entangling}, they performed the experiment of the HOM interference 
between two photons independently prepared  by SPDC 
with a cw pump light, and observed a visibility of 0.77. 
However, in order to perform various kinds of applications, 
such as quantum repeaters~\cite{yuan2008experimental, RevModPhys.83.33}, quantum relays~\cite{PhysRevA.66.052307, collins2005quantum}, measurement-device-independent quantum key distribution~(QKD)~\cite{PhysRevLett.108.130503, pirandola2015high, Scherer:11},
and distributed quantum computation~\cite{PhysRevA.89.022317}, a higher visibility will be desirable.
In this paper, we report on a demonstration of such a high-visibility HOM interference 
between two photons in the cw-pumped regime with high-resolution photon detectors. 
In the experiment, we introduced an extended method to observe 
an intrinsic value of the interference visibility. 
We observed a visibility of 0.87 $\pm$ 0.04, 
which clearly exceeds the previous result~\cite{halder2007entangling}. 
In addition, we derived a simple relationship between the visibility of the HOM interference and 
the second order intensity correlation function of the input light and background photons, 
and confirmed that the experimental results are in a good agreement with the relationship. 

\begin{figure}[t]
 \begin{center}
   \scalebox{0.38}{\includegraphics{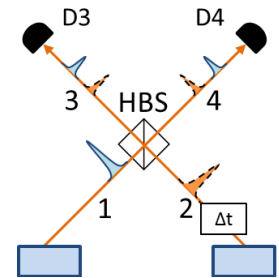}}
  \caption{(Color online)~
     The schematic diagram of the HOM interference using independent photon sources. 
     The relative delay is adjusted by changing the optical path length in mode~2 in 
     the case of the pulse pumped heralded single photons. 
     In the case of the cw pumped heralded single photons, 
     the relative delay is adjusted by changing the detection timings of the heralding photons.
     \label{fig:2pi}}
 \end{center}
\end{figure}
\begin{figure}[t]
 \begin{center}
   \scalebox{0.4}{\includegraphics{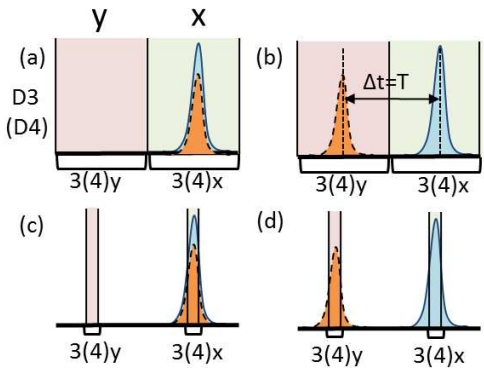}}
  \caption{(Color online)~The configurations for the coincidence time windows. 
     Each time window is divided into the two temporal modes $x$ and $y$. 
     The solid and dashed pulses correspond to the light pulses coming from mode~1 and mode~2, respectively. 
     (a,b)~Each of the time windows are
     set to be much larger than the pulse duration. 
     Two incoming pulses at D3~(D4) are shown with time delay $T$~(a) and without time delay~(b). 
     (c,d)~Each of the time windows are set to be shorter than the pulse duration for suppressing stray photon detection. 
     Two incoming pulses at D3~(D4) are shown with time delay $T$~(c) and without time delay~(d). 
     \label{fig:window}}
 \end{center}
\end{figure}

\section{Method}
We first review the standard method of time-resolved coincidence detection for the HOM interference experiments. 
Then we discuss an extended method that reduces an effect of stray photons or accidental coincidences 
by shortening a coincidence time window even below the coherence time of signal photons. 

Let us explain the standard method to observe the HOM interference 
of two independent light pulses. 
As shown in Fig.~\ref{fig:2pi}, 
the light pulses in modes 1 and 2 are mixed by a half beamsplitter~(HBS) 
followed by photon detectors D3 and D4. 
We assume that a relative delay~(timing difference) $\Delta t$ between two incoming light pulses is 
precisely determined in advance. The delay line is introduced in mode 2 
for the adjustment of $\Delta t$. 

As is well known, 
by counting the two-fold coincidence events between D3 and D4 
with detection time windows much larger than the pulse duration 
for various values of $\Delta t$, 
the HOM dip will be 
observed, while the single counts of D3 and D4 take constant values~\cite{PhysRevA.67.022301, PhysRevLett.96.110501, PhysRevA.79.040302, PhysRevA.81.021801, Tanida:12,
mcmillan2013two, Tsujimoto:15}. 
The visibility of the HOM interference is represented by $V=1-P_0/P_\infty$, 
where $P_0$ and $P_\infty$ are the two-fold coincidence probabilities with $\Delta t=0$ 
and $\Delta t=T$, respectively, where $T$ is chosen to be much larger than the pulse duration of input light. 

The two-fold coincidence probability 
$P$ can be understood as follows: As shown in Fig.~\ref{fig:window}, 
we assume that a common fixed time window of width $2T$ is adopted for D3 and D4, 
regardless of the amount of delay $\Delta t$. 
 Let us divide the window in two and denote the two halves by $x$ and $y$.
Figs.~\ref{fig:window}~(a) and (b) show the sketches of the light pulses 
with $\Delta t=0$ 
and that with $\Delta t=T$, respectively. 
The light from mode 1 is always detected at the center of $x$, 
while the one from mode 2 is detected in $x$ for $\Delta t=0$ and in $y$ for $\Delta t=T$. 
Let us denote by $3x$ a detection event at D3 in period $x$, which may be caused by the input pulse, 
stray background photons, or dark counting, and  define $4x$, $3y$, $4y$ similarly. 
Then, the coincidence probability $P$ determined in this conventional setup corresponds to the rate of occurrence of event 
($3x$ or $3y$) and ($4x$ or $4y$). 
An underlying assumption in this method is that the contribution of backgrounds should 
be the same for Figs.~\ref{fig:window}~(a) and \ref{fig:window}~(b), which is satisfied if the statistical properties of stray photons are time-invariant.

If background photons 
are not negligible, shortening the widths of time windows $x$ and $y$ 
is favorable. 
In this case, each width of the time window 
may be close to or even much shorter than the signal pulse width~\cite{halder2007entangling, 1367-2630-10-2-023027, PhysRevA.80.042321}, as shown 
in Fig.~\ref{fig:window}~(c) and (d). 
In this paper, we adopt this regime for suppressing stray photons. 
The coincidence probability $P$ is calculated from the rate of occurrence of event 
($3x$ or $3y$) and ($4x$ or $4y$), just as in the  conventional setup. 
The contribution of stray photons should be the same for 
Figs.~\ref{fig:window}~(c) and \ref{fig:window}~(d) if the statistical properties of stray photons are time-invariant, 
which is also normally assumed in the conventional setup. 

\begin{figure}[t]
 \begin{center}
  \includegraphics[width=\columnwidth]{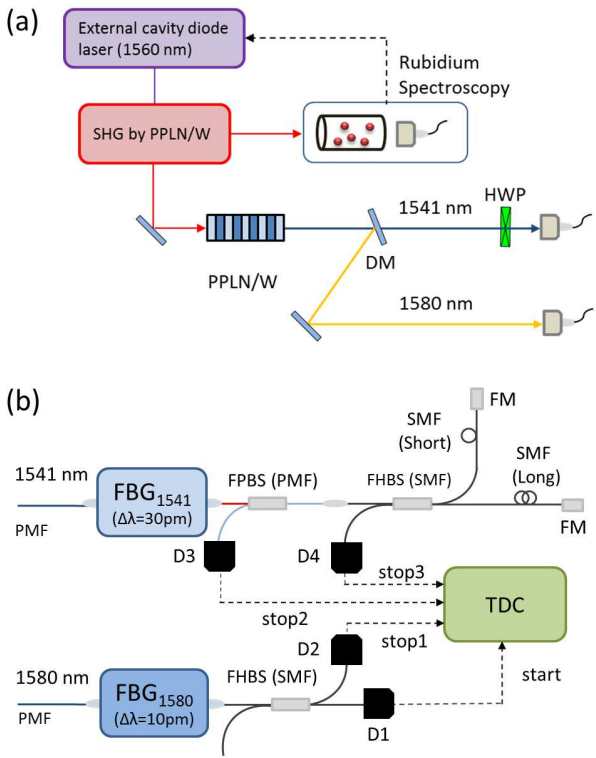}
  \caption{(Color online)~(a) The experimental setup of a photon pair source. The cw pump beam at 780~nm is obtained by
the second-harmonic generation based on a periodically-poled lithium niobate waveguide~(PPLN/W) pumped by an external
cavity diode laser working at 1560~nm. Non-degenerate photon pairs are generated by SPDC by another PPLN/W. (b)~The
experimental setup of the HOM interference. We performed the experiment of the HOM interference between photons at
1541~nm which are heralded by the photon detection at D1 and D4 with a time difference of $\Delta t$.
     \label{fig:HOMsetup}}
 \end{center}
\end{figure}

\section{Experiment}
\subsection{Photon pair source}
Our experimental setup is 
shown in Fig.~\ref{fig:HOMsetup}~(a). 
An initial beam from an external cavity diode laser~(ECDL) working at 1560~nm with a linewidth of 1.8~kHz is 
frequency doubled by using a periodically-poled lithium niobate waveguide~(PPLN/W)~\cite{Nishikawa:09}. 
The frequency of the ECDL is stabilized by the saturated 
absorption spectroscopy of rubidium atoms. 
The obtained cw light at 780~nm is set to be vertically~(V) polarized and is coupled to a 40-mm-long and 
type-0 quasi-phase-matched PPLN/W. 
It generates non-degenerate photon pairs at 1541~nm and 1580~nm by SPDC. 
The V-polarized photons at 1541~nm and 1580~nm 
are separated into different spatial modes by a dichroic mirror~(DM). 
The photons at 1541~nm are flipped to horizontal~(H) polarization by a half-wave plate~(HWP), 
and they are coupled to a polarization maintaining fiber~(PMF) 
followed by a fiber-based Bragg grating~(FBG). The photons at 1580 nm are also coupled to a PMF followed by 
a FBG. 

We first characterized our photon pair source by 
measuring the single counts and the coincidence counts just after the FBGs in Fig.~\ref{fig:HOMsetup}~(b). 
In order to evaluate the photon pair generation efficiency, we set the bandwidths of $\rm{FBG}_{1541}$ and $\rm{FBG}_{1580}$ to be 1~nm. 
For preventing the saturation of the detectors, we set the pump power 
to be $p=105~\rm{\mu}W$. The observed coincidence count rate is $C=1.8\times 10^5$~counts/(s$\cdot$nm)
and single count rates are $S_{1541}=2.0\times 10^6$~counts/(s$\cdot$nm) for 1541~nm and $S_{1580}=3.0\times 10^6$~counts/(s$\cdot$nm) for 1580~nm. 
The single count rates and the coincidence count rate are represented by 
$S_{1541}=\gamma p \eta_{1541}$, $S_{1580}=\gamma p \eta_{1580}$ and $C=\gamma p \eta_{1541}\eta_{1580}$, where 
$\gamma$ is a photon pair generation efficiency and $\eta_\lambda$ is a overall transimittance of the system 
for the photon at $\lambda$ nm. 
We estimated $\gamma=S_{1541}S_{1580}/(Cp)$ just after the PPLN/W 
to be about $3.2\times10^8$ pairs/(s$\cdot$mW$\cdot$nm). 
When we performed the HOM experiment described in detail later, we connected the output of the FBGs to the fiber-based optical circuit and set the bandwidths of $\rm{FBG}_{1541}$ to be 30~pm and $\rm{FBG}_{1580}$ to be 10~pm, respectively. 
In addition, we set the detection window to be 80~ps and the pump power to be 2.5~mW.  
In this setup, the observed coincidence count rate between detectors D1 and D3 was $1.0\times10^3$ counts/s.  

\subsection{Detection system} 
\begin{figure}[t]
 \begin{center}
\includegraphics[width=\columnwidth]{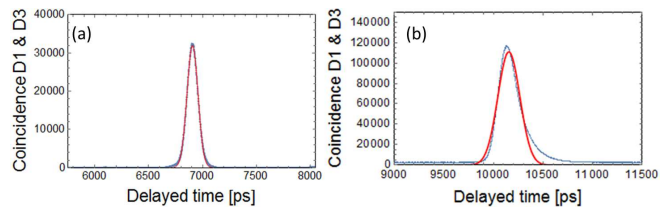}
  \caption{(Color online)~
    The relations between two-fold coincidence counts and detection timings. 
    The red solid curves are the Gaussian fit to the obtained data. 
    (a)~The result of the measurement of the timing jitter. 
    The pump power coupled to PPLN/W is set to be 3.5~$\mu$W, and 
    we directly connected the output of the FBGs~(3~nm) to D1 $\&$ D2. 
    The other combinations of detectors (D1, D3), (D1, D4) and (D2, D3) 
    are almost the same as (a). 
        (b)~The measurement result of the coherence time $\tau$. 
    The pump power coupled to PPLN/W is set to be 2.5~mW, and 
    we used $\rm{FBG}_{1580}$(10~pm) and $\rm{FBG}_{1541}$(30~pm). 
      \label{fig:Timingjitter}}
 \end{center}
\end{figure}

Next, we measured the timing jitter of the overall timing measurement system with superconducting nanowire single-photon detectors~(SNSPDs)~\cite{Miki:13}. 
For the measurement of the timing jitter, we used bandwidths~(3~nm) for $\rm{FBG}_{1541}$ and $\rm{FBG}_{1580}$ 
which correspond to the temporal width of 1.2~ps, 
and measured the arrival time of the picosecond photons by directly connecting the output of the FBGs to the detectors. 
In this setup, since the temporal width of the photons is sufficiently shorter than the timing jitter of the system, 
the distribution of arrival time difference of the photons reflects the timing jitter of the system. 
The typical experimental result is shown in Fig.~\ref{fig:Timingjitter}~(a). By fitting the experimental data by Gaussian, 
the deconvolved timing jitter is estimated to be 85~ps~FWHM for all of the four detectors. 
We also estimated the coherence time $\tau$, 
which is defined by the temporal width of the photon after $\rm{FBG}_{1541}$(30~pm) 
heralded by the half of the photon pairs which is filtered by $\rm{FBG}_{1580}$(10~pm). 
In our experiment, the temporal distribution obtained by the coincidence detection reflects 
the convolution of $\tau$ and timing jitters of the photon detectors.
Fig.~\ref{fig:Timingjitter}~(b) shows the two-fold coincidence count between D1 and D3, and the fitted Gaussian curve. 
By subtracting the effect of the timing jitter of each detector from the FWHM of the fitted Gaussian, 
$\tau$ is estimated to be 231~ps~FWHM. 

\subsection{HOM interference}
The HOM interference is performed by using a fiber-based optical circuit as shown in Fig.~\ref{fig:HOMsetup}~(b). 
The photons at 1580~nm are 
filtered by $\rm{FBG}_{1580}$ with a bandwidth of 10~pm.
After the FBG, the photons are split into two different spatial modes by a fiber-based half beamsplitter~(FHBS) 
followed by SNSPDs. 
The photon detections by D1 and D2 with a time difference of $\Delta t$ heralds two photons at 1541~nm with the same time difference $\Delta t$. 
They are filtered by $\rm{FBG}_{1541}$ with a bandwidth of 30~pm as shown in Fig.~\ref{fig:HOMsetup}~(b). 
After passing through a fiber-based polarizing beamsplitter~(FPBS), the two H-polarized photons propagate in a single-mode fiber~(SMF), and 
they are separated into a long path and a short path by a FHBS. 
The photons are reflected by a Faraday mirror~(FM) at the end of each path, at which the polarizations of the photons are flipped. 
After going back to the FHBS, the two photons are split into two paths and they are detected by D3 and D4.
We note that each of the two photons receives polarization fluctuation in the SMF. 
In the optical circuit, such fluctuations are much slower than the round-trip time of the photon propagation in each path 
and thus the birefringence effect in the SMF is automatically compensated after passing through each path by using the FM~\cite{muller1997plug}. 

\begin{figure}[t]
 \begin{center}
\includegraphics[width=\columnwidth]{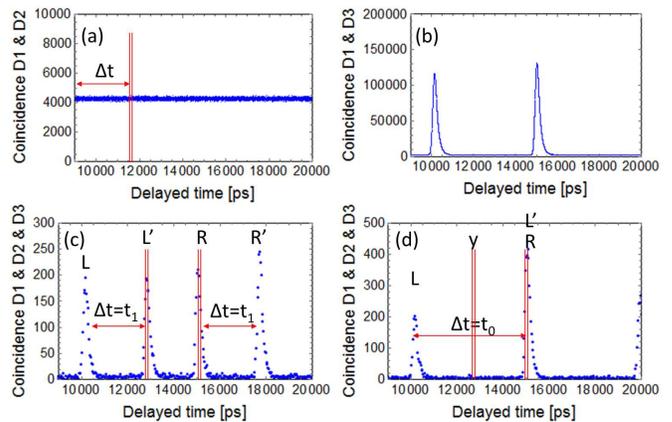}
  \caption{(Color online)~
   The coincidence counts between (a)~D1 \& D2 and (b)~D1 \& D3, 
   respectively. 
  (c)~The three-fold coincidence among D1, D2 and D3 
  for the start signal of D1. 
   The delay time of the stop signal of D2 is chosen to be $\Delta t=2.6$~ns. 
  (d)~The three-fold coincidence among D1, D2 and D3 
  for the start signal of D1. The delay time s chosen such that $\Delta t$ is equal to
  the time lag between photons coming from long and short paths.
  During the experiment, the pump power coupled to PPLN/W is set to be 2.5~mW. 
  \label{fig:coincidence}}
 \end{center}
\end{figure}

We collect the four-fold coincidence events by using a time-digital converter~(TDC). The electric signal from D1 is used as a start signal, and the electric 
signals from D2, D3 and D4 are used as stop signals. 
When the photons are detected at D1, the histograms of the delayed coincidence counts in the stop signals 
at D2 and D3 are obtained as shown in Figs.~\ref{fig:coincidence}~(a) and (b), respectively. 
In Fig.~\ref{fig:coincidence}~(b), 
left peak L and right peak R indicate the events where the photons at 1541~nm passed through the short 
and long paths, respectively. 
When we postselect the events where the stop signal of D2 has a delay $\Delta t$, additional two peaks L' and R' appear in the histogram of D3 
as in Fig.~\ref{fig:coincidence}~(c), which shows the three-fold coincidence among D1, D2 and D3. 
The two-fold coincidence between D1 and D4, 
and the three-fold coincidence among D1, D2 and D4 
show almost the same histgrams as Fig.~\ref{fig:coincidence}~(b) and (c), respectively. 
We chose a value of  $\Delta t=t_1$ such that the two peaks L' and R are separated, 
and defined the timing of the peak L' as $y$ and that of peak R as $x$. Then we determined 
the four-fold coincidence count $C_{\infty}$ from the events where the stop signals of D2, D3 and D4 
were recorded at the timings $t_1$, ($x$ or $y$) and ($x$ or $y$), respectively, with all 
of the coincidence windows having a width of 80~ps. 
This corresponds to the situation shown in Fig. ~\ref{fig:window}~(d). If we chose the delay for D2 to be $\Delta t=t_0$ such that peak L' is overlapped on R, 
the post-selected histogram of the three-fold coincidence among D1, D2 and D3 becomes as shown in Fig.~\ref{fig:coincidence}~(d). 
In this case, we determined the four-fold coincidence count $C_{0}$ by selecting the stop signals with the timing of $t_0$ for D2, 
and (x or y) for D3 and D4. 
This corresponds to the situation shown in Fig.~\ref{fig:window}~(c). 
The visibility is obtained by $V=1-C_{0}/C_{\infty}$. 

We set the pump power coupled to PPLN/W to be 2.5~mW. 
The total measurement time is 156 hours. 
By the experimental data, we obtained $C_{\infty}$ and $C_{0}$ as 87 counts and 11 counts, respectively. 
The observed visibility was 0.87 $\pm$ 0.04. This value 
is much higher than 0.77 obtained in Ref.~\cite{halder2007entangling}. 

\section{Discussion}
We consider the reason for the degradation of 
the observed visibility of the HOM interference. 
In our experiment, since the effect of the timing jitters of the detectors is 
estimated to be small due to the large coherence time, 
we only consider the effect of stray photons including multiple pair emission. 
Below we derive a simple relationship among the visibility and the intensity 
autocorrelation functions of the signal light pulses and stray photons. 
For simplicity, we assume that the signal photons and stray photons observed in a time window 
$x$ or $y$ are in a single mode. 
We define creation operators of the input and 
the output light of the HBS as $\hat{a}^{\dagger}_{ik}$ and $\hat{b}^{\dagger}_{jk}$, respectively, 
where $i=1,2$, $j=3,4$ and $k=x,y$. 
The above operators satisfy the
commutation relations 
$[\hat{a}_{ik}, \hat{a}^\dagger_{i'k'}]=\delta_{ii'}\delta_{kk'}$ and 
$[\hat{b}_{jk}, \hat{b}^\dagger_{j'k'}]=\delta_{jj'}\delta_{kk'}$. 
As in our experiment, we assume that the transmittance $\eta_i$ of the system including the 
detection efficiency of $\U{D}{i}$ for $i=3,4$ 
is much less than 1 
such that the events where two or more photons are simultaneously detected at the 
single detector are negligible, and the detection probabilities are proportional to the photon number 
in the detected mode.
The two-fold coincidence probability $P$ is expressed as 
$P=\eta_3\eta_4\expect{:(\hat{n}_{3x}+\hat{n}_{3y})(\hat{n}_{4x}+\hat{n}_{4y}):}$,  where 
$\hat{n}_{jk}=\hat{b}^\dagger_{jk}\hat{b}_{jk}$ is the number operator for output modes $j=3,4$ and $k=x,y$. 
$P_0$ and $P_\infty$ are calculated by using $P$ for the input signal light pulses 
coming from $1x$ \& $2x$ and $1x$ \& $2y$, respectively. 
As is often the case with practical settings, we assume that 
the signal light pulses and stray photons have no phase correlation 
and are statistically independent. 
We characterize the property of stray photons alone by the average photon 
number $n$ and the normalized intensity correlation function $g_n^{(2)}$. That is to say, if mode $ik$ does not include the signal light, 
$\expect{\hat{a}^\dagger_{ik}\hat{a}_{ik}}=n$ and $\expect{:(\hat{a}^\dagger_{ik}\hat{a}_{ik})^2:}/n^2 =g_n^{(2)}$ holds. Similarly, we define $s$ and $g_s^{(2)}$ such that, when mode $ik$ includes the signal light as well as stray photons, $\expect{\hat{a}^\dagger_{ik}\hat{a}_{ik}}=s$ and $\expect{:(\hat{a}^\dagger_{ik}\hat{a}_{ik})^2:}/s^2 =g_s^{(2)}$ holds.
From the definitions, we obtain 
$P_0=\eta_3\eta_4(s^2 g_s^{(2)}+n^2 g_n^{(2)}+4sn)/2$ and 
$P_\infty=\eta_3\eta_4(s^2 g_s^{(2)}+n^2 g_n^{(2)}+ (s+n)^2)/2$. 
Here we used a unitary operator $\hat{U}$ of the HBS satisfying 
$\hat{U}\hat{b}^\dagger_{3k}\hat{U}^\dagger
=(\hat{a}^\dagger_{1k}+\hat{a}^\dagger_{2k})/\sqrt{2}$ and 
$\hat{U}\hat{b}^\dagger_{4k}\hat{U}^\dagger
=(\hat{a}^\dagger_{1k}-\hat{a}^\dagger_{2k})/\sqrt{2}$. 
Thus $V$ is represented by 
\begin{eqnarray}
 V&=&\frac{(1-\chi)^2}{g_s^{(2)}+\chi^2g_n^{(2)}+(1+\chi)^2}, 
 \label{HOMvis}
\end{eqnarray}
where $\chi=n/s$. 
When $n\ll s$, 
the visibility is approximated by $V=1/(1+g_s^{(2)})$, 
and it takes maximum of $V=1$ for the input of genuine single photons~\cite{ikuta2016heralded}. 

In our experimental setup in Fig.~\ref{fig:HOMsetup}, the intensity correlation $g_s^{(2)}$ of the signal light could be measured by using the coincidence between D3 and D4 under the heralding of D1, if we run an additional experiment with the short arm detached from the FHBS. Instead, we calculated the same quantity from the same experimental data gathered for the main result.  This implies that the determined value $g_{ex}^{(2)}$ includes the contribution of the stray photons coming into FHBS from the short arm.
The single count probability $S_{3(4)}$ at D3(4) and coincidence count probability $C_{34}$ between D3 \& D4 
conditioned on the photon detection at D1 are represented by $S_{i}=\eta_i (s+n)/2$ 
and $C_{34}=\eta_3\eta_4 (s^2g_s^{(2)}+n^2g_n^{(2)})/4$.
The measured quantity is then given by 
$g_{ex}^{(2)}\equiv C_{34}/(S_3S_4)=(g_{s}^{(2)}+\chi^2g_{n}^{(2)})/(1+\chi)^2$, 
which is equal to $g_{s}^{(2)}$ for $\chi=0$. Using this measured quantity, the visibility in Eq.~(\ref{HOMvis}) is represented by
\begin{equation}
V=\frac{(1-\chi)^2}{(1+\chi)^2(1+g_{ex}^{(2)})}\label{HOMvis2}. 
\end{equation}
So far, we have assumed that each detector  receives only a single mode per window for simplicity. Rather surprisingly, the relation (\ref{HOMvis2}) still holds when detectors also receives stray photons in other modes, as long as the statistical independence among different modes is fulfilled (see Supplemental Material).
From the experimental results of 
$g^{(2)}_{ex}=5.3\times10^{-2}$ and $\chi=2.8\times10^{-2}$, 
we obtain $V=0.85$, which is in good agreement 
with the observed visibility of 0.87 $\pm$ 0.04 within a margin of error.  
These results indicate that the degradation of the visibility is 
mainly caused by stray photons. 
Since the main cause of the stray photons is  the accidental emission of photon pairs from SPDC, 
a higher visibility will be obtained by using a lower pump power 
and shorter timing selection for the photon detection. 

\section{Conclusion}
In conclusion, we have performed an experiment of the HOM interference between two photons produced by the 
two independent SPDC processes with cw pump light. 
The observed visibility is 0.87 $\pm$ 0.04, which 
is much higher than 0.77 observed in Ref.~\cite{halder2007entangling}. 
We also presented a simple relation 
between the visibility and the second order intensity correlation function,
and showed that it holds with a good approximation in this experiment. 
The relation is
convenient for the estimation of the possible visibility from the 
second order 
intensity correlation functions of the various input light sources. 
The results presented here will be useful for many applications such as QKD and quantum repeaters without 
the necessity of performing active synchronizations of the photon sources.

\begin{acknowledgements}
This work was supported by JSPS Grant-in-Aid for Scientific Research (A)~JP16H02214, (B)~JP25286077, (B)~JP26286068 and (B)~JP15H03704. RI, TY and NI were supported by JSPS Bilateral Open Partnership Joint Research Projects. YT was 
supported by JSPS Grant-in-Aid for JSPS Research Fellow JP16J05093. 
\end{acknowledgements}

\bibliographystyle{h-physrev}

\section*{Supplemental Material}
\renewcommand{\thefigure}{S\arabic{figure}}
\renewcommand{\theequation}{S\arabic{equation}}
\setcounter{figure}{0}
\setcounter{equation}{0}
\begin{figure}[h]
 \begin{center}
\includegraphics[width=\columnwidth]{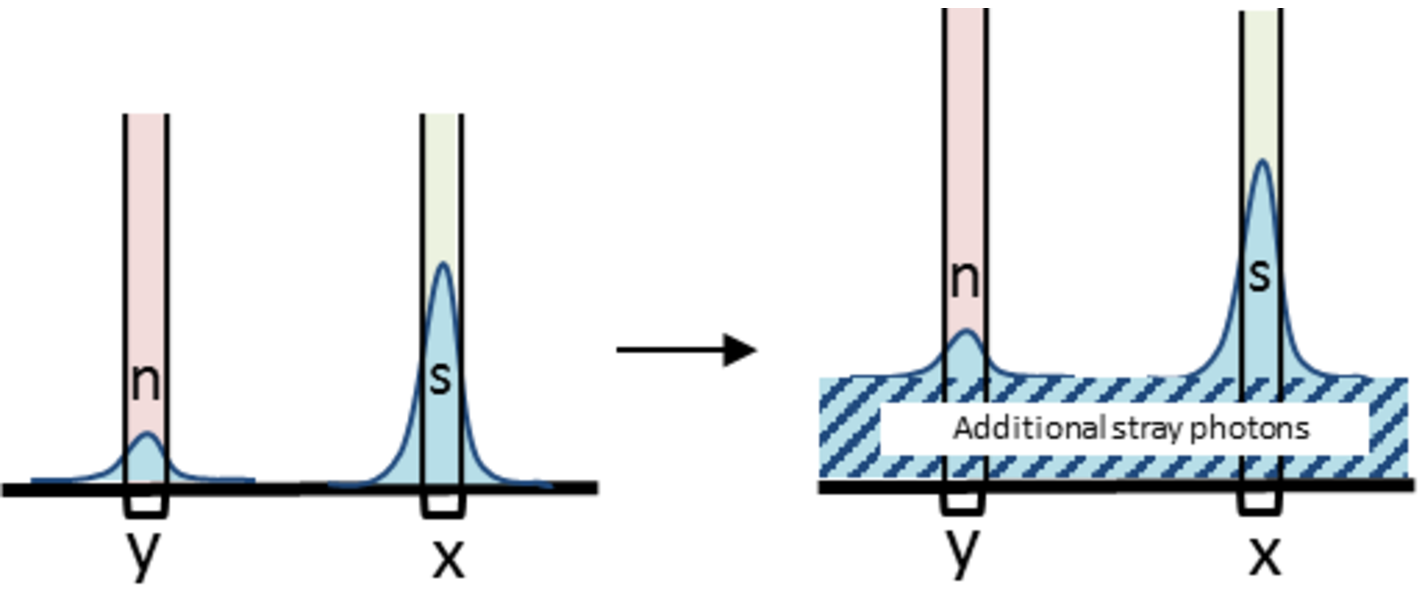}
  \caption{(Color online)~
     The left figure shows the photon distribution considered in the main text. The right figure presents the 
     sketch in the case where  additional stray photons in other modes are detected. 
  \label{fig:pulse}}
 \end{center}
\end{figure}

We show that the relation among $V$, $\chi$ and $g^{(2)}_{ex}$ 
in the main text holds true even under the presence of stationary photon noise in modes different 
from the signal modes, as long as noise photons in different modes are statistically independent. 
The situation is shown in Fig.~\ref{fig:pulse}. 
The additional noises accompanying the input signal pulse of mode $1x$ are decomposed into 
mutually orthogonal modes labeled by index $l$, and its creation operator is denoted by 
$\hat{a}^\dagger_{l,1x}$. The signal mode is orthogonal to every noise mode, 
namely, $[\hat{a}^\dagger_{1x}, \hat{a}^\dagger_{l,1x}]=0$. 
We further define the creation operators $\hat{a}^\dagger_{l,2x}$, $\hat{b}^\dagger_{l,3x}$, 
and $\hat{b}^\dagger_{l,4x}$ such that the four modes with a common index $l$ form the input and output modes of HBS. 
We define the modes in window $y$ similarly as $\hat{a}^\dagger_{l,iy}$, $\hat{b}^\dagger_{l,jy}$. 
Let us define sums of the photon number operators by  
$\hat{N}_{ik}=\sum_l\hat{a}^{\dagger}_{l,ik}\hat{a}_{l,ik}$ and 
$\hat{N}_{jk}=\sum_l\hat{b}^{\dagger}_{l,jk}\hat{b}_{l,jk}$. 
Since the noise photons are assumed to be stationary, we have 
$\expect{\hat{N}_{ix}} = \expect{\hat{N}_{iy}}$ and 
$\normal{\hat{N}_{3x}\hat{N}_{4x}} =\normal{\hat{N}_{3y} \hat{N}_{4y}}$. 
Define 
$N:=(\expect{\hat{N}_{1x}}+\expect{\hat{N}_{2x}})/2=(\expect{\hat{N}_{1y}}+\expect{\hat{N}_{2y}})/2$.

We first calculate the two-fold coincidence probability $P=\eta_3\eta_4\normal{(\hat{n}'_{3x}+\hat{n}'_{3y})(\hat{n}'_{4x}+\hat{n}'_{4y})}$ 
between D3 and D4 in Fig.~1 in the main text, where 
$\hat{n}'_{jk}=\hat{n}_{jk}+\hat{N}_{jk}$.   
When the input signal light pulses come from $1x$ \& $2x$, one of the four terms in $P$ is calculated as  
$\normal{\hat{n}'_{3x}\hat{n}'_{4x}}=\normal{(\hat{n}_{3x}+\hat{N}_{3x})(\hat{n}_{4x}+\hat{N}_{4x})}
=(\normal{\hat{n}^2_{1x}}+\normal{\hat{n}^2_{2x}}+2\expect{\hat{n}_{1x}+\hat{n}_{2x}}\expect{\hat{N}_{1x}+\hat{N}_{2x}})/4
+\normal{\hat{N}_{3x}\hat{N}_{4x}}
=s^2 g_{s}^{(2)}/2+2sN+\normal{\hat{N}_{3x}\hat{N}_{4x}}$, 
where $\expect{\hat{n}_{1x}}=\expect{\hat{n}_{2x}}=s$
and $\normal{\hat{n}^2_{1x}}=\normal{\hat{n}^2_{2x}}=s^2 g_{s}^{(2)}$. 
Here we used 
$\expect{\hat{n}_{ik}\hat{N}_{ik}}=\expect{\hat{n}_{ik}}\expect{\hat{N}_{ik}}$ 
from the assumption, and 
a unitary transformation by the HBS as 
$\hat{U}\hat{b}^{\dagger}_{l,3k}\hat{U}^\dagger
=(\hat{a}^{\dagger}_{l,1k}+\hat{a}^{\dagger}_{l,2k})/\sqrt{2}$ and 
$\hat{U}\hat{b}^{\dagger}_{l,4k}\hat{U}^\dagger
=(\hat{a}^{\dagger}_{l,1k}-\hat{a}^{\dagger}_{l,2k})/\sqrt{2}$
in addition to the transformation of the signal modes.  
Similarly, $\normal{\hat{n}_{3y}'\hat{n}_{4y}'}$ is calculated as $n^2 g_{n}^{(2)}/2+2nN+\normal{\hat{N}_{3x}\hat{N}_{4x}}$
, where $\expect{\hat{n}_{1y}}=\expect{\hat{n}_{2y}}=n$
and $\normal{\hat{n}^2_{1y}}=\normal{\hat{n}^2_{2y}}=n^2 g_{n}^{(2)}$. 
$\expect{\hat{n}_{3x}'\hat{n}_{4y}'}$ and $\expect{\hat{n}_{4x}'\hat{n}_{3y}'}$ are calculated as 
$sn+N(s+n)+N^2$. 
As a result, 
\begin{eqnarray}
P_0=&&\eta_3\eta_4((s^2 g_{s}^{(2)}+n^2 g_{n}^{(2)})/2+2N^2+4N(s+n)\nonumber\\
&&+2sn+2\normal{\hat{N}_{3x}\hat{N}_{4x}}).\label{S1} 
\end{eqnarray}
On the other hand, when the input signal light pulses come from $1x$ \& $2y$, 
$\normal{\hat{n}_{3x}'\hat{n}_{4x}'}$ and $\normal{\hat{n}_{3y}'\hat{n}_{4y}'}$ are calculated 
as $(s^2 g_{s}^{(2)}+n^2 g_{n}^{(2)})/4+N(s+n)+\normal{\hat{N}_{3x}\hat{N}_{4x}}$. 
Similarly, $\expect{\hat{n}_{3x}'\hat{n}_{4y}'}$ and $\expect{\hat{n}_{3y}'\hat{n}_{4x}'}$ are calculated as 
$(s+n)^2/4+N(s+n)+N^2$. As a result, we obtain 
\begin{eqnarray}
P_\infty=&&\eta_3\eta_4((s^2 g_{s}^{(2)}+n^2 g_{n}^{(2)})/2+2N^2+4N(s+n)\nonumber\\
&&+(s+n)^2/2+2\normal{\hat{N}_{3x}\hat{N}_{4x}}). \label{S2}
\end{eqnarray}

Next, we introduce the relation among $g_{ex}^{(2)}$, $g_{s}^{(2)}$ and $g_{n}^{(2)}$ 
in our experimental setup in Fig.~3 in the main text. 
The single count $S_{3(4)}$ at D3(4) and coincidence count $C_{34}$ between D3 \& D4 
conditioned on the photon detection at D1 are described by $S_{3(4)}=\eta_{3(4)}(s+n+2N)/2$, 
and $C_{34}=\eta_3\eta_4(s^2g_{s}^{(2)}+n^2g_{n}^{(2)}+4N(s+n)+4\normal{\hat{N}_{3x}\hat{N}_{4x}})/4$. 
The observed correlation function is expressed as 
\begin{eqnarray}
g_{ex}^{(2)}&=&\frac{C_{34}}{S_3S_4}\nonumber\\
&=&\frac{s^2g_{s}^{(2)}+n^2g_{n}^{(2)}+4N(s+n)+4\normal{\hat{N}_{3x}\hat{N}_{4x}}}{(s+n+2N)^2}. \label{S3}
\end{eqnarray}
By combining equation (\ref{S1}), (\ref{S2}) and (\ref{S3}), the visibility $V= 1-P_0/P_\infty$ is described by 
\begin{equation}
V=\frac{(1-\chi)^2}{(1+\chi)^2(1+g_{ex}^{(2)})}\label{HOMvis2}, 
\end{equation}
where $\chi= (n+N)/(s+N)$. 
\end{document}